\documentclass[12pt]{article}
\usepackage{graphicx,latexsym}

 \def\pd{\partial} \def\pp{\prime} \def\half{\frac{1}{2}} \def\fr{\frac} 

\def\a{\alpha} \def\b{\beta} \def\dl{\delta} \def\s{\sigma} \def\vphi{\varphi}  
 \def\lam{\lambda} \def\Lam{\Lambda} \def\gm{\gamma} \def\Gm{\Gamma}
  \def\nb{\nabla} \def\sq{\sqrt} 

\def\hg{{\hat g}}    
   \def\hG{{\hat G}} 

\def\bnb{{\bar \nabla}} \def\bg{{\bar g}} \def\bDelta{{\bar \Delta}} 
\def\bR{{\bar R}}  \def\bG{{\bar G}}

\def\barb{{\bar \beta}}

\begin{document}

\begin{titlepage}

\begin{flushright}
April 2017
\end{flushright}

\vspace{5mm}

\begin{center}
{\large {\bf Physical Cosmological Constant in Asymptotically Background-Free Quantum Gravity}}
\end{center}

\vspace{5mm}

\begin{center}
{\sc Ken-ji Hamada{}$^{a,b}$ and Mikoto Matsuda{}$^b$}
\end{center}

\begin{center}
{}$^a$ {\it Institute of Particle and Nuclear Studies, KEK, Tsukuba 305-0801, Japan} \\
{}$^b$ {\it Department of Particle and Nuclear Physics, SOKENDAI (The Graduate University for Advanced Studies), Tsukuba 305-0801, Japan}
\end{center}

\begin{abstract}
We study the effective potential in renormalizable quantum gravity with a single dimensionless conformal coupling without a Landau pole. In order to describe a background-free dynamics at the Planck scale and beyond, the conformal-factor field is quantized exactly in a nonperturbative manner. Since this field does not receive renormalization, the field-independent constant in the effective potential becomes itself invariant under the renormalization group flow. That is to say, it gives the physical cosmological constant. We explicitly calculate the physical cosmological constant at the one-loop level in the Landau gauge. We find that it is given by a function of renormalized quantities of the cosmological constant, the Planck mass and the coupling constant, and it should be the observed value. It will give a new perspective on the cosmological constant problem free from an ultraviolet cutoff.
\end{abstract}

\vspace{5mm}

\end{titlepage}

\section{Introduction}
\setcounter{equation}{0}

The gravitational theories on the basis of the Einstein action defined by the Ricci scalar cannot go beyond the Planck scale. So, the Planck mass scale gives an ultraviolet (UV) cutoff of such classical or quantum theories. Also, the existence of the UV cutoff is one of the reasons behind the cosmological constant problem \cite{weinberg89}. In order to resolve the problem, we have to build a field theory without the UV cutoff, which necessarily becomes quantum gravity.

The quantization of gravity based on the Einstein action \cite{ud, dewitt, veltman} has a lot of difficulties. First of all, the theory is not renormalizable. Furthermore, the path integral becomes unstable even in a nonperturbative manner, because the Ricci scalar is not bounded from below, like a scalar theory with an odd potential. In addition, we cannot eliminate a spacetime configuration with a singularity because the action becomes finite for such a spacetime solution.

In order to resolve these problems, quantum gravity theories involving the square of the Riemann curvature tensor have been proposed in the 1970s \cite{stelle, tomboulis, ft, ab}. In those days, however, one could not avoid the problem that the negative-metric mode emerges as a gauge-invariant mode as far as one dealt with all gravitational fields perturbatively.\footnote{ 
There is an idea on unitarity \cite{tomboulis} based on the work of Lee and Wick \cite{lw}, but it does not work in the UV limit where coupling constants vanish. }

To further resolve the ghost problem, we constructed renormalizable quantum conformal gravity several years ago \cite{hamada02,hamada14QG, hm} by applying a nonperturbative method learned from the development of two-dimensional quantum gravity \cite{polyakov, kpz, dk, david, seiberg, teschner}. The conformal factor of the metric field is treated exactly without introducing a coupling constant for it \cite{riegert, am, amm92, amm97, hs, hh, hamada05}, and as a result the theory has Becchi-Rouet-Stora-Tyutin (BRST) conformal symmetry in the UV limit, which represents the background-free property of quantum gravity as a gauge equivalency under conformal transformations \cite{hamada12M4, hamada12RxS3}. This symmetry makes ghost modes unphysical exactly. Physical states are then given by diffeomorphism-invariant real scalars only, which is consistent with scalar-dominated scale-invariant spectra of the early Universe \cite{wmap, planck}.

In renormalizable quantum gravity, the cosmological constant in the action, in general, receives renomalization so that it becomes running. Since a physically measurable quantity must be renormalization group (RG) invariant, it cannot be thought of as a physical cosmological constant. So, what is a RG-invariant cosmological constant has been one of the important problems in quantum gravity.

In this paper, we consider what the physical cosmological constant is in our renormalizable quantum gravity. Recall that, when we define a physical mass of a particle in quantum field theories, we usually adopt the on-shell renormalization scheme. However, such a scheme is not known for certain for renormalization of the cosmological constant in quantum gravity. In order to answer it, we here consider the effective potential with respect to the cosmological term that depends on the conformal-factor field only and discuss its RG structure. Owing to the nonrenormalization property of the conformal-factor field, we find that the field-independent constant in the effective potential becomes itself invariant under the RG flow, and thus it gives the physical cosmological constant.

We then calculate the physical cosmological constant explicitly at the one-loop level. The calculation is carried out in the Landau gauge in order to reduce the number of Feynman diagrams considerably and also to avoid some indeterminate factors. We show that the physical cosmological constant is given by a function of renormalized quantities of the cosmological constant, the Planck mass, and the dimensionless coupling constant, just like the pole mass can be written in terms of the renormalized mass and coupling constant.

\section{Brief Summary of Renormalizable Quantum Gravity}
\setcounter{equation}{0}

First, we briefly summarize recent developments of the renormalizable quantum field theory of conformal gravity with adding the Einstein-Hilbert action and the cosmological constant term \cite{hamada14QG, hm}. The theory has been formulated using dimensional regularization. The advantages of employing this regularization are that it preserves diffeomorphism invariance and also the theory becomes independent of the choice of the path integral measure owing to $\int d^D p = \dl^D (0) =0$. And also, there are no quadratic and quartic divergences, which are substantial in UV complete theories. On the other hand, the contributions from the measure such as conformal anomalies \cite{cd, ddi, duff, ds} are hidden between $D$ and four dimensions, and thus we have to determine the $D$ dependence of the action exactly.

\subsection{Quantum gravity action}

When we generalize fourth-order gravitational actions to $D$-dimensional ones, a lot of ambiguities emerge, unlike the case of ordinary gauge field action. In order to settle such ambiguities, we have recently analyzed Hathrell's RG equations for quantum field theories with conformal couplings in curved space \cite{hathrellS, hathrellQED, freeman}. We then found that the ambiguities disappear and the $D$ dependence of the gravitational action can be determined at all orders \cite{hamada14CS, hm}. The renormalizable quantum gravity has been defined by using this action, because it should reduce to the curved theory in the classical limit of gravity such as the large number limit of matter fields.

The quantum gravity action determined in this way has been expressed as
\begin{eqnarray*}
   S = \int d^D x \sqrt{g} \left[ \frac{1}{t_0^2} C_{\mu\nu\lam\s}^2  + b_0 G_D  - \frac{M_0^2}{2} R + \Lambda_0 + {\cal L}_{\rm matter} 
       \right] .
\end{eqnarray*}
The first term is the Weyl action given by the square of $D$-dimensional Weyl tensor $C_{\mu\nu\lam\s}$ as 
\begin{eqnarray*}
      C_{\mu\nu\lam\s}^2 = R_{\mu\nu\lam\s}^2 - \frac{4}{D-2} R_{\mu\nu}^2 + \frac{2}{(D-1)(D-2)} R^2 
\end{eqnarray*}
and the second term is the Euler density generalized to $D$ dimensions that is significant for the conformal-factor dynamics, which is exactly determined by solving Hathrell's RG equations as
\begin{eqnarray}
   G_D = G_4 + (D-4) \chi(D) H^2 ,
    \label{expression of G_D}
\end{eqnarray}
where $G_4= R_{\mu\nu\lam\s}^2 - 4 R_{\mu\nu}^2 + R^2$ is the usual Euler combination and $H=R/(D-1)$. The function $\chi(D)$ is expanded about four dimensions as $\chi(D) = \sum_{n=1}^{\infty} \chi_n (D-4)^{n-1}$, whose coefficient $\chi_n$ can be determined order by order. The first two terms have been calculated explicitly as $\chi_1 =1/2$ and $\chi_2=3/4$ \cite{hamada14CS, hm}.\footnote{ 
Furthermore, from the analysis of QED in curved space \cite{hamada14CS}, it has been found that $\chi_3$ is given by $1/3$, which becomes necessary in calculations of three loops or more.} 
We have then shown that these are universal values independent of the gauge group and the contents of matter fields as far as they are conformally coupled.\footnote{ 
If there is a nonconformal dimensionless coupling, we have to add the pure $R^2$ term to the action in addition to $C_{\mu\nu\lam\s}^2$ and $G_D$.}

The bare quantity $t_0$ is a single dimensionless gravitational coupling constant, while $b_0$ is not an independent coupling as mentioned below. The bare mass parameters $M_0$ and $\Lam_0$ are the Planck mass and the cosmological constant, respectively. The last term ${\cal L}_{\rm matter}$ denotes conventional second-order matter fields with dimensionless conformal couplings.

The conformal anomaly associated with the action (\ref{expression of G_D}) is then expressed in the form $E_D = G_D -4\chi(D) \nb^2 H$ \cite{hamada14CS}. Here, it is significant that the familiar ambiguous $\nb^2 R$ term is fixed completely, and due to $\chi_1=1/2$ this combination reduces at $D \to 4$ to $E_4=G_4 -2 \nb^2 R/3$ proposed by Riegert \cite{riegert}.

\subsection{Renormalization procedure and asymptotic background freedom}
 
The perturbation in $t_0$ implies that the metric field is expanded about a conformally flat spacetime satisfying $C_{\mu\nu\lam\s}=0$, which is defined by
\begin{eqnarray}
    g_{\mu\nu}  =  e^{2\phi}\bg_{\mu\nu},  \quad
    \bg_{\mu\nu} = (\hg e^{t_0 h_0})_{\mu\nu} = \hg_{\mu\lam} \left( \dl^\lam_{\nu} + t_0 h^\lam_{0\,\nu} 
                       + \fr{t_0^2}{2} h^\lam_{0\,\s} h^\s_{0\,\nu} + \cdots \right) ,
            \label{expansion of metric field}
\end{eqnarray}
where $h_{0\mu\nu}= \hg_{\mu\lam}h^\lam_{0\nu}$, $h^\mu_{0\mu}=0$, and $\hg_{\mu\nu}$ is the background metric. The quantum gravity can be thus described as a quantum field theory defined on the background $\hg_{\mu\nu}$.

The significant feature of this theory is that the conformal factor $e^{2\phi}$ is written in the exponential form to maintain its positivity and treated exactly without introducing its own coupling constant. It ensures the independence under the conformal change of the background $\hg_{\mu\nu} \to  e^{2\s}\hg_{\mu\nu}$, because, as is apparent from (\ref{expansion of metric field}), this change can be absorbed by rewriting the integration variable as $\phi \to \phi - \s$ in quantized gravity, which is algebraically represented as BRST conformal symmetry. Consequently, we can choose the flat background without affecting the results.

The renormalization factors for the traceless tensor field and the coupling constant are defined as usual by $h_{0\mu\nu}=Z_h^{1/2}h_{\mu\nu}$ and $t_0 = \mu^{2-D/2} Z_t t$, where $\mu$ is an arbitrary mass scale to make up for the loss of mass dimensions, and thus the renormalized coupling $t$ becomes dimensionless. On the other hand, since we treat the conformal-factor field exactly without introducing the corresponding coupling constant, diffeomorphism invariance requires that it does not receive renormalization such that\footnote{ 
This nonrenormalization theorem has been demonstrated explicitly in loop calculations of higher order \cite{hamada02, hamada14QG, hm}.} 
\begin{eqnarray}
    Z_{\phi}=1 .
        \label{Z_phi = 1}       
\end{eqnarray}
It can be easily understand from the fact that gauge invariance results in the relationship between the renormalization factor of the coupling constant and that of the corresponding field.\footnote{ 
For instance, $Z_e Z_3^{1/2}=1$ in QED such that $e_0 A_{0\mu}= e A_\mu$ at $D=4$. In general, the argument precisely holds only for the background gauge field in the background field method \cite{abbott}. For the $\phi$ field, however, it is true because this field is not gauge fixed, unlike the traceless tensor field, so that the renormalization factor of $\phi$ is the same to that of its background field and it becomes unity from diffeomorphism invariance. This fact is used in the next section.}
No coupling constant thus implies that there is no field-renormalization factor. This is one of the most significant properties in our renormalization calculations, which reflects the independence of how to choose the background metric as mentioned above.

The beta function of $\a_t =t^2/4\pi$ is defined by $\b_t \equiv (\mu/\a_t) d \a_t/d\mu = D-4 + \barb_t$. At the one-loop level, we obtain
\begin{eqnarray*}
   \barb_t = - \left[ \fr{1}{120} \left( N_S + 6 N_F + 12 N_A \right) + \fr{197}{30} \right] \fr{\a_t}{4\pi} 
\end{eqnarray*}
for $N_S$ conformally coupled scalars, $N_F$ fermions, and $N_A$ gauge fields  \cite{cd, ddi, duff}. The last term is the contribution from the gravitational field \cite{ft, amm92, hs}. The coupling $\a_t$ thus indicates the asymptotic freedom, which justifies performing the perturbation theory about conformally flat spacetime.

Here, we emphasize that the asymptotic freedom of the traceless tensor field does not mean the realization of a picture in which free gravitons are propagating in the flat spacetime, because the conformal factor is still nonperturbative and spacetime totally fluctuates quantum mechanically. So, we call it the asymptotic background freedom. We can thus go beyond the Planck scale.

In addition to these renormalization factors, we also introduce the bare parameter $b_0$ to renormalize UV divergences proportional to the $G_D$ term. Since its volume integral becomes topological at four dimensions, $b_0$ is not an independent dynamical coupling, and thus we expand it in a pure-pole series as
\begin{eqnarray}
  b_0 = \fr{\mu^{D-4}}{(4\pi)^{D/2}} \sum_{n=1}^\infty \fr{b_n}{(D-4)^n} . 
     \label{definition of bare coefficient b_0}
\end{eqnarray}
Since the expansion of the volume integral of $G_D$ starts from $o(D-4)$, the finite terms, namely, Wess-Zumino actions for conformal anomalies, come out with offsetting this zero by the pole in $b_0$, and those describe the dynamics of the conformal-factor field. Here, $b_n ~(n \geq 2)$ depends on the coupling constants only, while the simple-pole residue has a coupling-independent part, and thus it is divided as 
\begin{eqnarray}
    b_1 = b + b_1^\pp , 
       \label{definition for prime b_1}
\end{eqnarray}
where $b_1^\pp$ is coupling dependent and $b$ is a constant part.

In order to carry out the renormalization systematically incorporating the conformal-factor dynamics induced quantum mechanically, we have proposed the following procedure. For the moment, $b$ is regarded as a new coupling constant. The effective action is then finite up to the topological term as follows:
\begin{eqnarray*}
    \Gm = \fr{\mu^{D-4}}{(4\pi)^{D/2}} \fr{b-b_c}{D-4} \int d^D x \sq{\hg} \hG_4 + \Gm_{\rm ren}(\alpha_t,b) ,
\end{eqnarray*} 
where $\Gm_{\rm ren}$ is the finite part obtained by the standard renormalization procedure. The divergent term exists in a curved background only. The one-loop constant $b_c$ can be calculated independently of $b$ \cite{cd, ddi, duff, ft, amm92, hs}, which is given by
\begin{eqnarray}
    b_c = \fr{1}{360} \left( N_S + 11 N_F + 62 N_A \right) + \fr{769}{180}  .
       \label{value of coefficient b_c}
\end{eqnarray}
After the renormalization is carried out, we take $b=b_c$ at four dimensions. In this way, we obtain the effective action whose dynamics is governed by a single dimensionless gravitational coupling $\alpha_t$.

From the RG equation $\mu db_0/d\mu=0$, we obtain the following expression:
\begin{eqnarray}
   \mu \fr{db}{d\mu} = (D-4) \barb_b  ,
      \label{RG equation of b}
\end{eqnarray}
where $\barb_b$ is a finite function given by 
\begin{eqnarray*}
    \barb_b = - \left( \fr{\pd b_1}{\pd b} \right)^{-1} 
        \left( b_1  + \a_t \fr{\pd b_1}{\pd \a_t} \right).
\end{eqnarray*}
Here, in order to be able to replace the coupling $b$ with the constant $b_c$ at the end, the condition $\mu db/d\mu = 0$ should be satisfied at four dimensions. Therefore, (\ref{RG equation of b}) ensures the validity of the renormalization procedure proposed above.

From the RG analysis of QED and QCD in curved space, we find that $b_1^\pp$ in (\ref{definition for prime b_1}) arises at the fourth order of the gauge-coupling constant \cite{hathrellQED, freeman, hamada14CS, hm}. From this fact and the similarity between the gauge field and the traceless tensor field, we assume that the $\a_t$ dependence of $b_1^\pp$ is also given by $b_1^\pp = o(\a_t^2)$ and thus $\barb_b = -b +o(\a_t^2)$.\footnote{ 
This assumption should be verified through explicit two-loop calculations of three-point functions of the traceless tensor field or indirect calculations using the RG equation, but this work is so hard and has not been completed yet.}

\subsection{Propagators and interactions}

In the following, we take the flat background $\hg_{\mu\nu} =\dl_{\mu\nu}$. The Weyl action in $D$ dimensions is then expanded as follows:
\begin{eqnarray}
      \fr{1}{t_0^2} \int d^D x \hbox{$\sq{g}$} F_D
         &=& \fr{1}{t_0^2} \int d^D x  e^{(D-4)\phi}{\bar C}_{\mu\nu\lam\s}^2 
              \nonumber \\
         &=& \int d^D x \left\{ \fr{1}{t_0^2} {\bar C}_{\mu\nu\lam\s}^2
             + \fr{D-4}{t_0^2} \phi  \,  {\bar C}_{\mu\nu\lam\s}^2 + \cdots \right\} .
           \label{expansion of Weyl action}
\end{eqnarray}
Here, the gravitational quantities with the bar are defined by using the metric $\bg_{\mu\nu}$ in (\ref{expansion of metric field}). The first term of the right-hand side gives the propagator and self-interactions of the traceless tensor field. The second and other terms are the induced Wess-Zumino interactions related to the conformal anomaly.

The kinetic term of the traceless tensor field is given by
\begin{eqnarray*}
     \int d^D x \left\{  \fr{D - 3}{D - 2} \left( h_{0\mu\nu} \pd^4 h_{0\mu\nu}  +  2\chi_{0\mu} \pd^2 \chi_{0\mu} \right)
         -  \fr{D - 3}{D - 1} \chi_{0\mu} \pd_{\mu}\pd_{\nu}\chi_{0\nu}   \right\}
\end{eqnarray*}
where $\chi_{0\mu}=\pd_\nu h_{0\mu\nu}$ and $\pd^2 = \pd_\mu \pd_\mu$. The same lower indices denote contraction in the flat metric $\dl_{\mu\nu}$. According to the standard procedure of gauge fixing, we introduce the following gauge-fixing term \cite{ft}:
\begin{eqnarray*}
     S_{\rm gf}  =  \int d^D x  \left\{ \fr{1}{2\zeta_0}  \chi_{0\mu} N_{\mu\nu} \chi_{0\nu} \right\} ,
\end{eqnarray*}
where
\begin{eqnarray*}
      N_{\mu\nu}=  \fr{2(D-3)}{D-2} \biggl( 
                 -2 \pd^2\dl_{\mu\nu} + \frac{D-2}{D-1} \pd_{\mu}\pd_{\nu} \biggr)   .
\end{eqnarray*}
The gauge parameter is renormalized as $\zeta_0 = Z_h \zeta$. We here disregard the ghost action, because the ghost field is not coupled with the conformal-factor field directly so that it is not necessary in the following calculations.

Let us present the propagator of the traceless tensor field. The equation of motion is now given by $K^{(\zeta)}_{\mu\nu,\lam\s}(k) h_{\lam\s}(k)=0$ in momentum space, where
\begin{eqnarray}
    K^{(\zeta)}_{\mu\nu,\lam\s}(k)  &=&
         \fr{2(D \!-\! 3)}{D - 2} \biggl\{ I^{\rm H}_{\mu\nu,\lam\s} k^4 
         + \fr{1 \!-\! \zeta}{\zeta} \biggl[  \half k^2 \bigl( \dl_{\mu\lam} k_\nu k_\s  +  \dl_{\nu\lam} k_\mu k_\s 
                \nonumber \\
    &&  + \dl_{\mu\s} k_\nu k_\lam  +  \dl_{\nu\s} k_\mu k_\lam \bigr) 
        - \fr{1}{D - 1} k^2 \bigl( \dl_{\mu\nu} k_\lam k_\s + \dl_{\lam\s} k_\mu k_\nu \bigr) 
                \nonumber \\
    &&  + \fr{1}{D(D \!-\! 1)} \dl_{\mu\nu} \dl_{\lam\s} k^4  - \fr{D \!-\! 2}{D \!-\! 1} k_\mu k_\nu k_\lam k_\s 
         \biggr] \biggr\} 
           \label{K in arbitrary gauge}
\end{eqnarray}
and $I^{\rm H}_{\mu\nu, \lam\s} = ( \dl_{\mu\lam}\dl_{\nu\s}+ \dl_{\mu\s}\dl_{\nu\lam} )/2 - \dl_{\mu\nu}\dl_{\lam\s}/D$. By solving the inverse of $K^{(\zeta)}_{\mu\nu,\lam\s}$, we obtain the propagator in the arbitrary gauge as
\begin{eqnarray}
        \langle h_{\mu\nu}(k) h_{\lam\s}(-k) \rangle = \fr{D-2}{2(D-3)} \frac{1}{k^4} I^{(\zeta)}_{\mu\nu, \lam\s}(k) ,
          \label{propagator of traceless tensor field in arbitrary gauge}
\end{eqnarray}
where 
\begin{eqnarray}
   I^{(\zeta)}_{\mu\nu, \lam\s}(k)  &=&
         I^{\rm H}_{\mu\nu, \lam\s}  + (\zeta - 1) \Biggl[ \half \biggl( \dl_{\mu\lam} \fr{k_\nu k_\s}{k^2}  
         +  \dl_{\nu\s} \fr{k_\mu k_\lam}{k^2}  +  \dl_{\mu\s} \fr{k_\nu k_\lam}{k^2} 
              \nonumber \\
   &&    +  \dl_{\nu\lam} \fr{k_\mu k_\s}{k^2} \biggr)
         - \fr{1}{D-1} \left( \dl_{\mu\nu} \fr{k_\lam k_\s}{k^2} +  \dl_{\lam\s} \fr{k_\mu k_\nu}{k^2} \right)
          \nonumber \\
   &&    + \fr{1}{D(D-1)} \dl_{\mu\nu} \dl_{\lam\s} 
         - \fr{D-2}{D-1} \fr{k_\mu k_\nu k_\lam k_\s}{k^4}  
        \Biggr] .
           \label{I in arbitrary gauge}
\end{eqnarray}
This tensor satisfies 
\begin{eqnarray*}
   k_\mu I^{(\zeta)}_{\mu\nu, \lam\s}(k) 
   = \zeta \left( \half k_\lam \dl_{\nu\s} + \half k_\s \dl_{\nu\lam} - \fr{1}{D} k_\nu \dl_{\lam\s} \right) , 
\end{eqnarray*}
and thus it becomes transverse when $\zeta=0$. The choice of $\zeta=0$ is called the Landau gauge, while $\zeta =1$ is called the Feynman gauge.

The kinetic term and self-interaction terms of the conformal-factor field are derived from the $b_0 G_D$ action. From (\ref{expression of G_D}) and (\ref{definition of bare coefficient b_0}), this action is expanded as follows: 
\begin{eqnarray}
     &&  b_0  \int  d^D x \sq{g} G_D
         =  \fr{\mu^{D-4}}{(4\pi)^{D/2}} \int d^D x \Biggl\{
          \biggl(  \fr{b_1}{D - 4}+\fr{b_2}{(D - 4)^2} + \cdots \biggr) \bG_4
            \nonumber  \\
     && \quad
         + \biggl( b_1 +\fr{b_2}{D - 4} +\cdots \biggr)
         \biggl( 2 \phi \bDelta_4 \phi +\bG_4 \phi -\fr{2}{3} \bnb^2 \bR \phi + \fr{1}{18}\bR^2 \biggr)
             \nonumber   \\
     &&  \quad
         + \bigl[ (D - 4)b_1 +\cdots \bigr] \bigl(  \phi^2 \bDelta_4 \phi + 3 \phi \bnb^4 \phi + \cdots \bigr)
         + \cdots \Biggr\} ,
          \label{expansion of G_D action}
\end{eqnarray}
where $\sqrt{g} \Delta_4$ is the fourth-order differential operator that becomes conformally invariant at four dimensions defined by $\Delta_4 = \nb^4 + 2R^{\mu\nu}\nb_\mu \nb_\nu - 2R \nb^2/3 + \nb^\mu R \nb_\mu/3$ \cite{riegert}. The terms that we do not use in the next section are denoted by the dots here.

The first group of the expansion (\ref{expansion of G_D action}) gives the counterterm subtracting UV divergences proportional to the Euler term $\bG_4$, which determine the residue $b_n$ in (\ref{definition of bare coefficient b_0}). The second group gives the Riegert action \cite{riegert}, which is the Wess-Zumino action for the conformal anomaly $E_4$. It includes the bilinear term of the conformal-factor field as
\begin{eqnarray*}
    \fr{\mu^{D-4}}{(4\pi)^{D/2}} 2 b \int d^D x   \phi \pd^4 \phi 
\end{eqnarray*}
at the lowest of the perturbations. Since this term is independent of the coupling $t$, we can use it as the kinetic term, and the propagator is then given by 
\begin{eqnarray}
    \langle \phi(k) \phi(-k) \rangle = \mu^{4-D} \frac{(4\pi)^{D/2}}{4b} \frac{1}{k^4} .
       \label{propagator of conformal-factor field}
\end{eqnarray}
Therefore, quantum corrections from this field are expanded in $1/b$, which corresponds to considering the large-$N$ expansion for the number of matter fields (\ref{value of coefficient b_c}).  Since the conformal-factor field does not propagate at $b \to \infty$, it gives the classical limit of gravity.

The third group of (\ref{expansion of G_D action}) gives the self-interaction among the conformal-factor fields. Since it has the $D-4$ factor, it becomes effective at the one-loop level and more. And also, there are many interactions including the traceless tensor field, but most of them can be dropped here when we employ the Landau gauge.

In the following, all calculations are carried out in the Landau gauge in order to reduce the number of Feynman diagrams and also to obtain physically acceptable results directly. It is because in the arbitrary gauge the $b \bnb^2 \bar{R} \phi$ interaction in (\ref{expansion of G_D action}) becomes effective and then yields contributions with a positive power of $b$ that do not vanish in the classical limit \cite{hm}. We think that such an unphysical behavior will disappear at last, but it is difficult to show that explicitly at present.

\subsection{Renormalizations of mass parameters \cite{hamada14QG, hm}}

The Einstein-Hilbert action is expanded up to the second order of the coupling constant as
\begin{eqnarray*}
     -\fr{M_0^2}{2} \int d^D x \sq{g} R 
     &=& -\fr{M_0^2}{2} \int d^D x e^{(D-2)\phi} \left\{ \bar{R} -(D-1) \bnb^2 \phi \right\} 
               \nonumber \\
     &=& \fr{3}{2} M_0^2 \int d^D x e^{(D-2)\phi} \biggl\{
                   \fr{D-1}{3} \pd^2 \phi  + \fr{t_0^2}{12} \pd_\lam h_{0 \mu\nu} \pd_\lam h_{0 \mu\nu} 
                  + \cdots  \biggr\} ,
\end{eqnarray*}
where the dots denote the interaction terms that do not contribute to loop calculations using the Landau gauge in the next section. The renormalization factor is defined by $M_0^2 = \mu^{D-4} Z_{\rm EH} M^2$. The anomalous dimension for the Planck mass is then defined by $\gm_{\rm EH} \equiv - (\mu/M^2) d M^2/d\mu = D-4 + {\bar \gm}_{\rm EH}$, where $\bar{\gm}_{\rm EH}=\mu d (\log Z_{\rm EH})/d\mu$, which has been calculated in the Landau gauge as
\begin{eqnarray}
    {\bar \gm}_{\rm EH} = \fr{1}{b} + \fr{1}{b^2} + \fr{5}{4} \fr{\a_t}{4\pi} .
      \label{anomalous dimension of Planck mass}
\end{eqnarray} 
Here, $o(1/b^2)$ comes from two-loop diagrams, and others are from one-loop diagrams. At the end, $b$ is replaced with $b_c$ at four dimensions.

The cosmological term is simply written in terms of the exponential factor of the $\phi$ field as
\begin{eqnarray*}
    \Lam_0 \int d^D x \sq{g} = \Lam_0 \int d^D x  e^{D\phi} .
\end{eqnarray*}
The renormalization factor is defined by $\Lam_0 = \mu^{D-4} Z_\Lam ( \Lam + L_M M^4 )$, where $L_M$ is the pure-pole term. The anomalous dimension for the cosmological constant is defined by $\gm_\Lam \equiv - (\mu/\Lam) d \Lam/d\mu = D-4 + {\bar \gm}_\Lam + (M^4/\Lam) {\bar \dl}_\Lam$, where $\bar{\gm}_\Lam = \mu d (\log Z_\Lam)/d \mu$ and ${\bar \dl}_\Lam = \mu dL_M/d\mu -(D-4)L_M + ( {\bar \gm}_\Lam -2 {\bar \gm}_{\rm EH} ) L_M$. The calculation in the Landau gauge has been carried out up to the first order of $\a_t$ as
\begin{eqnarray}
     \bar{\gm}_\Lam = \fr{4}{b} + \fr{8}{b^2} + \fr{20}{b^3} - \fr{310}{9 \, b} \fr{\a_t}{4\pi}  , \qquad
        \bar{\dl}_\Lam = - \fr{9(4\pi)^2}{8 \, b^2}  
       \label{anomalous dimension of cosmological constant}
\end{eqnarray}
with $b=b_c$. Here, $o(1/b^3)$ in $\bar{\gm}_\Lam$ comes from three-loop diagrams, and $o(\a_t/b)$ is from two-loop diagrams. There is no correction of the first order of $\a_t$ to $\bar{\dl}_\Lam$. This anomalous dimension vanishes at $b \to \infty$, which is consistent with the classical limit of gravity.

Here, note that the $\a_t$-independent terms in $\bar{\gm}_{\rm EH}$ (\ref{anomalous dimension of Planck mass}) and $\bar{\gm}_\Lam$ (\ref{anomalous dimension of cosmological constant}) agree with the exact solutions of these anomalous dimensions derived using BRST conformal symmetry at $\a_t=0$ \cite{amm97, hh, hamada12M4, hamada12RxS3}.

\section{Physical Cosmological Constant in Quantum Gravity}
\setcounter{equation}{0}

Now, let us discuss the effective action in our renormalizable quantum gravity. We here consider the effective action that is expanded in powers of the conformal-factor field background $\sigma$ as
\begin{eqnarray*}  
       \Gamma (\sigma) &=& \sum_n \frac{1}{n!} \int d^{D} x_{1} \ldots d^{D} x_{n} 
                           \Gamma^{(n)}(x_{1}, \ldots, x_{n} ) \sigma(x_{1}) \ldots \sigma(x_{n})  
                           \\
        &=& \sum_n \frac{1}{n!} \int \frac{d^{D} k_{1}}{(2\pi)^{D}} \ldots \frac{d^{D} k_{n}}{(2\pi)^{D}} 
                           (2\pi)^D \delta^{(D)}(k_{1}+ \cdots + k_{n} )
                      \\
                 && \times  \Gamma^{(n)}(k_{1}, \ldots, k_{n} ) \sigma(k_{1}) \ldots \sigma(k_{n}) , 
\end{eqnarray*}
where $\Gamma^{(n)}$ is the renormalized $n$-point Green function given by the sum of all 1PI Feynman diagrams with $n$ external legs of $\sigma$.

The RG analysis of the 1PI Green function $\Gamma^{(n)}$ can be carried out as in the case of the $\varphi^4$ theory \cite{gl, symanzik, callan, thooft, weinberg73}. One of the crucial differences is that the conformal-factor field is not renormalized such that $Z_{\phi}=1$ and also for its background. Therefore, the renormalized $\Gamma^{(n)}$ is the same as the bare one, and thus $\mu d\Gamma^{(n)}/d\mu = 0$ is satisfied.\footnote{In the $\varphi^4$ theory, the field receives renormalization so that $\Gamma^{(n)}$ is not RG invariant, though $\Gamma(\vphi)$ itself is RG invariant.}

The effective potential $V$ is given by the zero momentum part of $\Gamma^{(n)}(k_1, \ldots, k_n)$, which is expressed as
\begin{eqnarray*}
     \Gm (\sigma)|_V = \int d^D x V(\s) = \sum_{n} \fr{1}{n!} \Gamma^{(n)}(0, \ldots, 0) \int d^{D} x \sigma^{n}(x) . 
\end{eqnarray*}
The diffeomorphism invariance implies that $\Gamma^{(n)}(0, \ldots, 0)  = v D^{n}$ and thus the effective potential has the form
\begin{eqnarray}
     V(\sigma)  = v e^{D\sigma(x)}  . 
       \label{form of effective potential}
\end{eqnarray}
The RG equation implies that $v$ is scale invariant such as
\begin{eqnarray*}
     \mu \fr{d}{d\mu} v = 0. 
\end{eqnarray*}
We thus find that the effective potential gives the physical cosmological constant, which can be observed cosmologically.

Before calculating the physical cosmological constant at the one-loop level explicitly, we first see the RG structure of the 1PI Green function, which will give a RG improvement of the effective action.

\subsection{RG structure}

The RG equation is derived from the condition $\mu d\Gm^{(n)}/d\mu =0$, which gives the following equation:
\begin{eqnarray}
   \left( \mu \frac{\partial}{\partial \mu} + \beta_{t} \alpha_{t} \frac{\partial}{\partial \alpha_{t}}
          - \gamma_{\Lambda} \Lambda \frac{\partial}{\partial \Lambda} - \gamma_{EH} M^{2} \frac{\partial}{\partial M^{2}} \right)
          \Gamma^{(n)} \left( k_{j}, \alpha_{t}, \Lambda, M^{2}, \mu \right) = 0,   \quad
        \label{RG equation 1}
\end{eqnarray}
where we take $D=4$ and thus the differential term $(D-4)\bar{\beta}_b \partial/\partial b$ is removed.

Changing the momentum variable as $k_j \to \lambda k_j$ and doing the dimensional analysis, we find that $\Gamma^{(n)}$ has the following form:
\begin{eqnarray*}
    \Gamma^{(n)} \left( \lambda k_{j}, \alpha_{t}, \Lambda, M^{2}, \mu \right) 
    = \mu^{4} \Omega^{(n)} \left( \frac{\lambda k_{j}}{\mu}, \alpha_{t}, \frac{\Lambda}{\mu^{4}}, \frac{M^{2}}{\mu^{2}} \right) .
\end{eqnarray*}
This implies that $\Gamma^{(n)}$ satisfies the differential equation
\begin{eqnarray}
   \left( \mu \frac{\partial}{\partial \mu} + 4 \Lambda \frac{\partial}{\partial \Lambda} + 2 M^{2} \frac{\partial}{\partial M^{2}} 
          + \lambda \frac{\partial}{\partial \lambda} - 4 \right)
          \Gamma^{(n)} \left( \lambda k_{j}, \alpha_{t}, \Lambda, M^{2}, \mu \right) = 0 . \quad
        \label{scale equation}
\end{eqnarray}
Therefore, combining (\ref{RG equation 1}) and (\ref{scale equation}) and removing the partial derivative of $\mu$, we obtain the expression
\begin{eqnarray}
   && \biggl( - \lambda \frac{\partial}{\partial \lambda} + \beta_{t} \bigl( \alpha_{t} \bigr) \alpha_{t} \frac{\partial}{\partial \alpha_{t}}
          - \Bigl[ 4 + \gamma_{\Lambda} \bigl( \alpha_{t}, \Lambda, M^{2} \bigr) \Bigr] \Lambda \frac{\partial}{\partial \Lambda} 
              \nonumber \\
   &&  - \Bigl[ 2 + \gamma_{EH} \bigl( \alpha_{t} \bigr)  \Bigr] M^{2} \frac{\partial}{\partial M^{2}}  + 4 \biggr)
          \Gamma^{(n)} \left( \lambda k_{j}, \alpha_{t}, \Lambda, M^{2}, \mu \right) = 0 .
        \label{RG equation 2}   
\end{eqnarray}

Here, we introduce the running coupling constant $\tilde{\alpha}_{t}(\lambda)$, the running cosmological constant $\tilde{\Lambda}(\lambda)$, and the running Planck mass $\tilde{M}(\lambda)$, which are defined by the following differential equations:
\begin{eqnarray}
   - \lambda \frac{d}{d \lambda} \tilde{\alpha}_{t} (\lambda) 
       &=& \beta_{t} \bigl( \tilde{\alpha}_{t}(\lambda) \bigr) \, \tilde{\alpha}_{t}(\lambda) ,
               \nonumber \\
   - \lambda \frac{d}{d \lambda} \tilde{\Lambda} (\lambda) 
       &=& - \Bigl[ 4 + \gamma_{\Lambda} \bigl( \tilde{\alpha}_{t}(\lambda), \tilde{\Lambda}(\lambda), \tilde{M}^{2}(\lambda) \bigr) \Bigr] 
       \tilde{\Lambda}(\lambda) ,
               \nonumber \\
   - \lambda \frac{d}{d \lambda} \tilde{M}^{2} (\lambda) 
       &=& - \Bigl[ 2 + \gamma_{EH} \bigl( \tilde{\alpha}_{t}(\lambda) \bigr) \Bigr] \tilde{M}^{2} (\lambda) .
         \label{defining equations of running couplings}
\end{eqnarray}
If we replace $\a_t$, $\Lam$, and $M^2$ in Eq. (\ref{RG equation 2}) with the corresponding running quantities $\tilde{\a}_t$, $\tilde{\Lam}$, and $\tilde{M}^2$, respectively, we find that this equation can be written with the help of the defining equations (\ref{defining equations of running couplings}) as
\begin{eqnarray*}
   \left( - \lambda \frac{d}{d \lambda} + 4 \right) 
   \Gamma^{(n)} \left( \lambda k_{j}, \tilde{\alpha}_{t}(\lambda), \tilde{\Lambda}(\lambda), \tilde{M}^{2}(\lambda), \mu \right) = 0 .
\end{eqnarray*}
Here, note that this RG equation is written in terms of the total differential with respect to $\lambda$, not the partial one. The solution is thus given by
\begin{eqnarray}
    \Gamma^{(n)} \left( \lambda k_{j}, \tilde{\alpha}_{t}(\lambda), \tilde{\Lambda}(\lambda), \tilde{M}^{2}(\lambda), \mu \right)
    = \lambda^{4} \Gamma^{(n)} \left( k_{j}, \alpha_{t}, \Lambda, M^{2}, \mu \right) 
      \label{solution of RG equation}
\end{eqnarray}
under the conditions of $\tilde{\alpha}_{t}(1)=\alpha_{t}$, $\tilde{\Lambda}(1)=\Lambda$, and $\tilde{M}^{2}(1)=M^{2}$.

From the solution (\ref{solution of RG equation}) with $k_j=0$ and the expression (\ref{form of effective potential}), we obtain the following equation:
\begin{eqnarray*}
    V \bigl( \tilde{\alpha}_{t}(\lambda), \tilde{\Lambda}(\lambda), \tilde{M}^{2}(\lambda), \mu \bigr) 
    = \lambda^{4} V \bigl( \alpha_{t}, \Lambda, M^{2}, \mu \bigr) . 
\end{eqnarray*}
Thus, the physical cosmological term improved by the RG equation is given by
\begin{eqnarray*}
    V = \tilde{v} (\lambda) e^{4\sigma(x)}  ,
\end{eqnarray*}
where 
\begin{eqnarray}
   \tilde{v}(\lambda) = \lambda^{-4} v( \tilde{\alpha}_{t}(\lambda), \tilde{\Lambda}(\lambda), \tilde{M}^{2}(\lambda), \mu)   
       \label{running cosmological constant}
\end{eqnarray}
is the physical cosmological constant, which does not depend on the RG parameter such that $d\tilde{v}(\lambda)/d\lambda = 0$.

\subsection{Explicit form of the physical cosmological constant}

As seen above, the effective potential gives the physical cosmological constant that is invariant under the RG flow. Let us here calculate the explicit form of it at the one-loop level, in which the background field $\sigma$ is taken to be a constant. We then consider the large $b$ limit, while the ratios $\Lambda/b$ and $M^4/b^2$ are taken to be the same order and also $\a_t/4\pi \sim 1/b$ is assumed. In this limit, the one-loop approximation becomes valid, and loop corrections to the effective potential are written by a function of these ratios.

The conformal-factor field is here divided into the constant background and quantum field $\vphi$ as
\begin{eqnarray*}
    \phi = \sigma + \vphi  .
\end{eqnarray*}
Expanding the gravitational action up to the second order of the quantum fields, $\vphi$ and $h_{\mu\nu}$, in the Landau gauge, we obtain the following action:
\begin{eqnarray*}
    S_{\rm kin} = S_\phi + S_h + S_{\rm c}  ,
\end{eqnarray*}
where each term is given by
\begin{eqnarray*}
    S_\phi &=& \int d^{D} x \mu^{D-4} \Biggl\{  
         \frac{1}{(4\pi)^{D/2}} \left[  2b \vphi \partial^{4} \vphi 
           + (D-4) b ( 2 \sigma + 3 ) \vphi \partial^{4} \vphi \right]
              \nonumber  \\
        && + \frac{(D-1)(D-2)}{2} M^{2} e^{(D-2)\sigma} \vphi \partial^{2} \vphi 
           +  \Lambda e^{D\sigma} \left( 1 + \frac{D^2}{2} \vphi^2 \right)  \Biggr\} ,
              \nonumber \\
    S_h &=& \int d^{D} x \biggl\{ 
          \frac{1}{2} h_{\alpha\beta} K^{(0)}_{\alpha\beta,\gamma\delta} h_{\gamma\delta} 
          + (D-4) \frac{D-3}{D-2} \sigma h_{\alpha\beta} \partial^{4} h_{\alpha\beta} 
              \nonumber  \\
        && - \frac{t^{2}}{8} M^{2} e^{(D-2)\sigma} h_{\alpha\beta} \partial^{2} h_{\alpha\beta} \biggr\} ,
                \\ 
    S_c &=& \int d^{D} x  \mu^{D-4} \left[ (Z_{\Lambda} -1 ) \Lambda + Z_{\Lambda} L_{M} M^{4} \right]  e^{D\sigma} .
\end{eqnarray*}
The kinetic term of the traceless tensor field, whose momentum representation is given by (\ref{K in arbitrary gauge}), is considered in the Landau gauge. The terms with the $D-4$ factor in $S_\phi$ and $S_h$ come from the induced Wess-Zumino interactions in (\ref{expansion of G_D action}) and (\ref{expansion of Weyl action}), respectively. In the Landau gauge, there are no contributions from the terms including curvature functions explicitly in (\ref{expansion of G_D action}). The renormalization factors in the last counterterm that is necessary for the one-loop calculation of effective potential are given by 
\begin{eqnarray*}
     Z_{\Lambda} -1 &=&  - \frac{2}{b} \left( - \frac{2}{D-4} - \gm + \log 4\pi \right) ,
          \nonumber \\
     L_{M} &=& \left( \frac{9(4\pi)^{2}}{16 b^{2}} - \frac{5}{64} \a_t^2  \right) \left( - \frac{2}{D-4} -\gm + \log 4\pi \right)
\end{eqnarray*}
in the modified minimal subtraction scheme. Here, note that the $o(\a_t^2)$ term of $L_M$ gives extra correction $5 \a_t^2/32$ to $\bar{\dl}_\Lam$ in (\ref{anomalous dimension of cosmological constant}), which is necessary in the approximation we are considering here.

\subsubsection{Loops of the conformal-factor field}

We first calculate the contribution from the conformal-factor field to the effective potential. In order to normalize the action, we rescale the quantum field $\vphi$ as
\begin{eqnarray*}
   \vphi = \sqrt{\frac{(4\pi)^{\frac{D}{2}}}{4b \mu^{D-4}[ 1+ (D-4)( \sigma + \frac{3}{2}) ]}} \psi .
\end{eqnarray*}
Apart from the classical term of $\Lam e^{D\s}$, we then obtain the following expression:
\begin{eqnarray*}
    S_\phi =  \frac{1}{2} \int \frac{d^{D} k}{(2\pi)^{D}} \psi(k) D_{\psi} \psi(-k) 
           =  \frac{1}{2} \int \frac{d^{D} k}{(2\pi)^{D}} \psi(k) \left[ k^4 - A k^2 + B \right] \psi(-k) ,   
\end{eqnarray*}
where $A$ and $B$ are defined by
\begin{eqnarray*}
   A &=& \frac{(4\pi)^{\frac{D}{2}}(D-1)(D-2)}{4b [1+ (D-4)(\sigma + \frac{3}{2})]} M^{2} e^{(D-2)\sigma} ,
      \nonumber \\
   B &=& \frac{(4\pi)^{\frac{D}{2}}D^{2}}{4b [1+ (D-4)(\sigma + \frac{3}{2})]} \Lambda e^{D\sigma}  . 
\end{eqnarray*}

\begin{figure}[h]
\begin{center}
\includegraphics[scale=0.7]{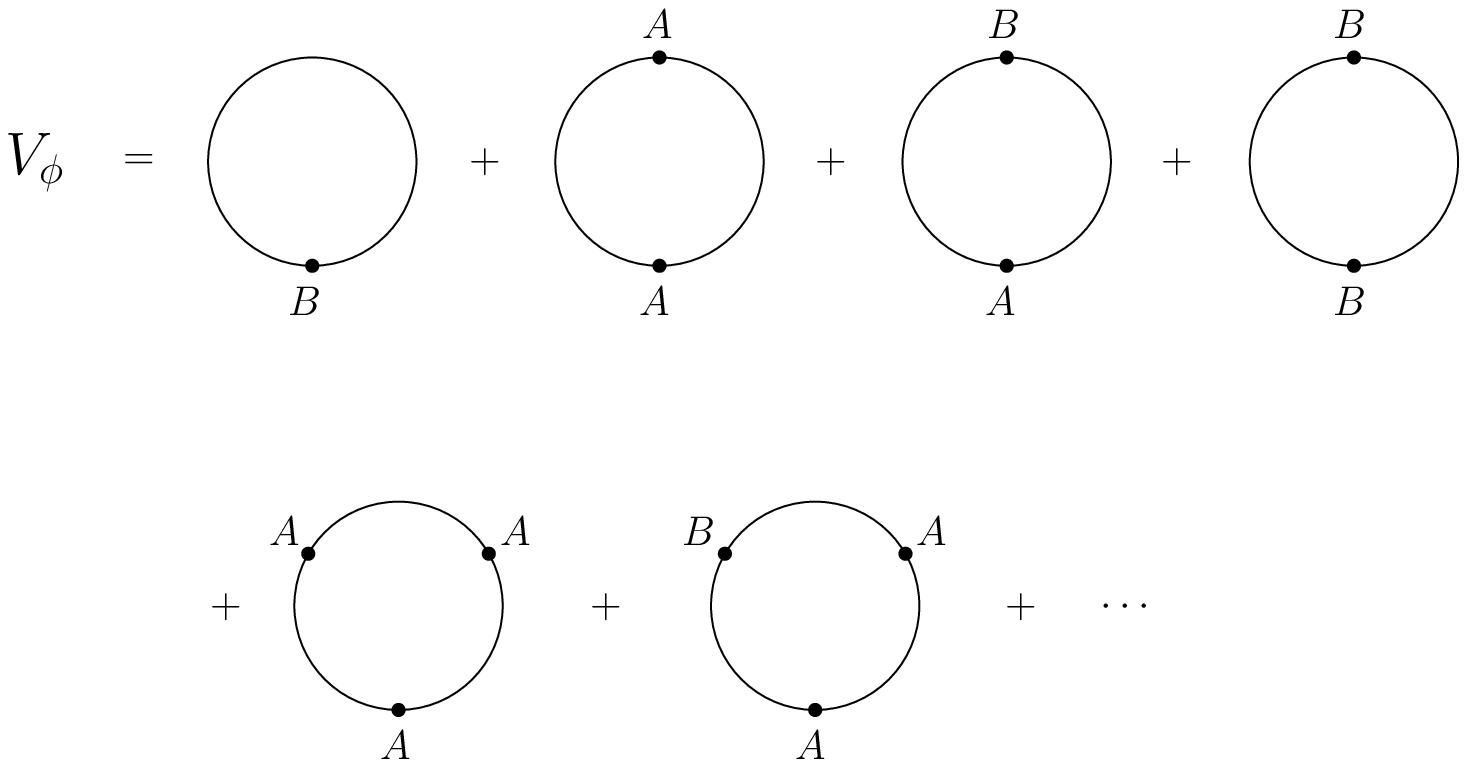}
\end{center}
\vspace{-6mm}
\caption{\label{effective potential for varphi}{\small One-loop contributions from the conformal-factor field.}}
\end{figure}

The one-loop correction to the effective potential is then expressed as follows:
\begin{eqnarray*}    
    V_{\phi} &=& - \log \left[\det ( D_{0}^{-1} D_{\psi} ) \right]^{-\frac{1}{2}}
            \nonumber  \\
       &=&  \frac{1}{2} \int \frac{d^{D} k}{(2\pi)^{D}} \log \left( 1 - \frac{A}{k^2} + \frac{B}{k^4} \right) ,
\end{eqnarray*}
where $D_{0}=k^{4}$ is the inverse of the propagator of the rescaled field $\psi$. The corresponding Feynman diagrams are depicted in Fig. \ref{effective potential for varphi}. Expanding the logarithmic function into power series of $A$ and $B$, we obtain the following expression:
\begin{eqnarray} 
     V_{\phi} &=& \frac{1}{2} \sum_{n=1}^{\infty} \frac{(-1)^{n-1}}{n}
          \int \frac{d^{D} k}{(2\pi)^{D}} \left( - \frac{A}{k^2} + \frac{B}{k^4} \right)^{n}
            \nonumber  \\
       &=&  \frac{1}{2} \sum_{n=1}^{\infty} \sum_{m=0}^{n} \frac{(-1)^{n-1}}{n} \frac{n!}{m!(n-m)!} (-A)^{m} B^{n-m} I_{2n-m}(z) ,  
          \label{V_varphi expression with integral formulas}
\end{eqnarray}
where the loop integral $I_{l}$ is defined by
\begin{eqnarray*}
     I_{l}(z) = \int \frac{d^{D} k}{(2\pi)^{D}} \frac{1}{(k^{2}+z^{2})^{l}}   
\end{eqnarray*}
and $z$ is an infinitesimal mass to regularize IR divergences. After carrying out the calculation, we take $z$ to be zero.

In the fourth-order quantum field theory, IR divergences become stronger than those of conventional second-order theories. So, throughout the loop calculations, we have to introduce such a fictitious small mass that violates diffeomorphism invariance. We will then see that all IR divergences indeed cancel out, including the consistency check, especially in the calculation of the effective potential in which more strong IR divergences arise.

The integral $I_{1}$ vanishes at the limit $z \to 0$, while $I_{2}$ has the UV and IR divergences as
\begin{eqnarray*}
    I_{2}(z) = \fr{\Gm(2-\fr{D}{2})(z^2)^{\fr{D}{2}-2}}{(4\pi)^{\fr{D}{2}}}  
       =  \frac{1}{(4\pi)^{2}} \left( - \frac{2}{D-4} - \gamma + \log 4\pi - \log z^{2} \right) .
\end{eqnarray*}
The integral with $l > 2$ has the IR divergence only written in powers of $z$ as
\begin{eqnarray*}
    I_{l}(z) =  \frac{1}{(4\pi)^{2}} \frac{1}{(l-1)(l-2)} \left( \frac{1}{z^{2}} \right)^{l-2} . 
\end{eqnarray*}
Substituting these integral values into (\ref{V_varphi expression with integral formulas}), we obtain the following expression:
\begin{eqnarray}
    V_{\phi} &=& \frac{1}{(4\pi)^2} \Biggl[ \left( \frac{B}{2} - \frac{A^{2}}{4} \right) 
        \left( - \frac{2}{D-4} - \gamma + \log 4\pi - \log z^{2} \right) 
        + \frac{AB}{4} \frac{1}{z^{2}} - \frac{B^2}{24} \frac{1}{z^{4}} \Biggr]
           \nonumber \\
     && + \frac{1}{2(4\pi)^2} \sum^{\infty}_{n=3} \sum^{n}_{m=0} \frac{(-1)^{n-1}}{n} \frac{n!}{m!(n-m)!} 
        \frac{(-1)^{m}A^{m} B^{n-m} (z^{2})^{2-2n+m}}{(2n-m-1)(2n-m-2)} .  
          \label{V_varphi expression with z} 
\end{eqnarray}
The sum of the infinite series part can be evaluated using the formula given in the Appendix. We can then take the limit of $z \to 0$ and show that IR divergences indeed cancel out. In this way, we obtain the expression that has UV divergences only as follows:
\begin{eqnarray*}
  V_{\phi} &=& \frac{1}{(4\pi)^2} \left( \frac{B}{2} - \frac{A^{2}}{4} \right) 
        \left( - \frac{2}{D-4} - \gamma + \log 4\pi  \right)
              \nonumber \\
       &&  + \frac{1}{(4\pi)^2} \Biggl[ \frac{1}{8} \bigl( 2B - A^{2} \bigr) (3 - \log B ) 
           - \frac{A}{4} \sqrt{4B-A^{2}} \arccos \left( \frac{A}{2\sqrt{B}} \right)
        \Biggr] .
\end{eqnarray*}

Substituting the expressions of $A$ and $B$, the part with the pole is now expanded as
\begin{eqnarray*}
    \frac{1}{(4\pi)^2} \left( \frac{B}{2} - \frac{A^2}{4} \right) \frac{-2}{D-4}
    &=& \mu^{D-4} e^{D\sigma} \Biggl[ 
        - \frac{2}{b} \Lambda \left( \frac{2}{D-4} - 2\sigma - 2 + \log 4\pi - \log \mu^2 \right)
                     \nonumber \\
    &&  + \frac{9 \pi^{2}}{b^2} M^4 \left( \frac{2}{D-4} - 2\sigma - \frac{8}{3} 
        + 2 \log 4\pi - \log \mu^{2} \right)   \Biggr] .
\end{eqnarray*}
The UV divergences are subtracted by the counterterm $S_c$. Here,  to fit the form of the counterterm, the $\mu$ dependence is recovered through the relation $\mu^{D-4}[1 - (D-4) \log \mu ] = 1 + o[(D-4)^2]$. Combining all finite terms and taking $D=4$ with replacing $b$ with $b_c$, we obtain the following effective potential:
\begin{eqnarray}
    V_{\phi} &=& e^{4\sigma} \Biggl[ \frac{\Lambda}{b_c} ( 7 - 2 \log 4\pi ) 
            - \frac{9\pi^{2} M^{4}}{2b_c^2} \left( \frac{25}{3} - 4 \log 4\pi \right) 
                \nonumber \\
         && - \left( \frac{\Lambda}{b_c} - \frac{9\pi^{2}M^{4}}{2b_c^{2}} \right) \log \frac{64\pi^{2}\Lambda}{b_c \mu^{4}}
            - \frac{6\pi M^{2}}{b_c} \sqrt{\frac{\Lambda}{b_c}- \frac{9\pi^{2}M^{4}}{4b_c^2}} 
              \arccos \left( \frac{3\pi M^{2}}{2\sqrt{b_c\Lambda}} \right) \Biggr].
                \nonumber \\
           \label{result of V_varphi}
\end{eqnarray}
Here, note that apart from the overall factor the $\sigma$ dependence disappears.

\subsubsection{Loops of the traceless tensor field}

\begin{figure}[h]
\begin{center}
\includegraphics[scale=0.7]{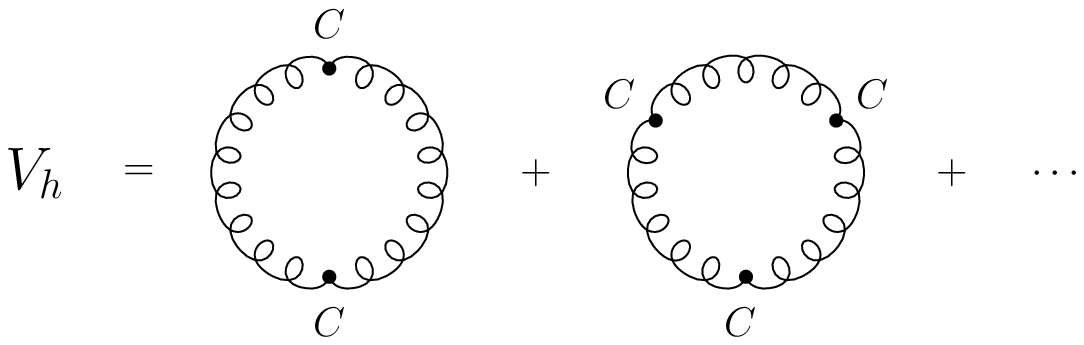}
\end{center}
\vspace{-6mm}
\caption{\label{effective potential for h}{\small One-loop contributions from the traceless tensor field.}}
\end{figure}

We next calculate the contribution from the traceless tensor field to the effective potential in Landau gauge, whose diagrams are depicted in Fig. \ref{effective potential for h}.\footnote{ 
There are diagrams, not depicted here, that are obtained by replacing one of $C$ with $(D-4)\s$. However, they could be absorbed in the definition of $C$ by changing the normalization of the field, as in the case of the conformal-factor field.} We here rewrite the action in the form
\begin{eqnarray*}  
     S_h =  \frac{1}{2} \int \frac{d^{D} k}{(2\pi)^D} h_{\mu\nu}(k) K^{h}_{\mu\nu,\lambda\sigma}(k) h_{\lambda\sigma}(-k) . 
\end{eqnarray*}
The momentum function $K^h_{\mu\nu,\lambda\sigma}$ is given by
\begin{eqnarray*}
     K^{h}_{\mu\nu,\lambda\sigma} &=& K^{(0)}_{\mu\nu, \alpha\beta} 
       \left\{ I^H_{\alpha\beta,\lambda\sigma} + \left[ (D-4) \sigma + \frac{C}{k^{2}} \right] 
       I^{(0)}_{\alpha\beta,\lambda\sigma} \right\}  ,
          \nonumber  \\ 
       C &=& t^{2} \frac{(D-2)}{8(D-3)} M^{2} e^{(D-2)\sigma}  ,
\end{eqnarray*}
where $K^{(0)}_{\mu\nu,\lambda\sigma} = K^{(\zeta)}_{\mu\nu,\lambda\sigma} |_{\zeta \to 0}$ and $I^{(0)}_{\mu\nu,\lambda\sigma}$ is (\ref{I in arbitrary gauge}) at $\zeta =0$. The one-loop correction to the effective potential is then given by
\begin{eqnarray*}   
   V_{h} &=& - \log \left[ \det \left( (K_{(0)}^{-1})_{\mu\nu,\alpha\beta} K^h_{\alpha\beta,\lambda\sigma} \right) \right]^{-\frac{1}{2}}
        \nonumber \\
     &=&   \frac{(D+1)(D-2)}{4} \sum_{n=1}^{\infty} \frac{(-1)^{n-1}}{n} \left[ 
             C^{n} I_{n}(z) + (D-4)n \sigma C^{n-1} I_{n-1}(z) \right], 
\end{eqnarray*}
where $(K_{(0)}^{-1})_{\mu\nu,\lambda\sigma}$ is the free propagator of the traceless tensor field (\ref{propagator of traceless tensor field in arbitrary gauge}) in the Landau gauge and we use $Tr[(I^{(0)})^{n}] = Tr[I^{(0)}]= (D+1)(D-2)/2$. Evaluating loop integrals as before, we obtain the following expression:
\begin{eqnarray}
    V_{h} &=& \frac{(D+1)(D-2)}{4(4\pi)^{2}} \biggl\{ \left[ \frac{1}{2} - (D-4) \sigma \right] C^{2} 
              \left( \frac{2}{D-4} - \gamma + \log 4\pi + \log z^{2} \right)
                 \nonumber \\
           && + z^{4} g \left( \frac{C}{z^{2}} \right) + (D-4) \sigma C^{2} h \left( \frac{C}{z^{2}} \right) \biggr\} ,  
             \label{V_h expression with z}
\end{eqnarray}
where the functions $g$ and $h$ are defined by 
\begin{eqnarray*}
   g(x) &=& \sum_{n=1}^\infty \fr{(-1)^{n-1}}{n(n+1)(n+2)} x^{n+2}
         = \fr{(1+x)^2}{2} \log (1+x) - \fr{3}{4} x(x+2) ,
            \nonumber \\
   h(x) &=& \sum_{n=1}^\infty \fr{(-1)^n}{n(n+1)} x^n
         = \fr{1+x}{x} \left[ 1 - \log (1+x) \right] . 
\end{eqnarray*}
The function (\ref{V_h expression with z}) also becomes finite in the $z \to 0$ limit, and thus all IR divergences cancel out. We then obtain the following expression: 
\begin{eqnarray}
   V_{h} &=& \frac{1}{(4\pi)^2} \Biggl[  
             - \frac{(D+1)(D-2)}{8} C^{2} \mu^{D-4} 
              \left( - \frac{2}{D-4} - \gamma + \log 4\pi + \log \mu^{2}  \right)
                \nonumber \\
         && - 5 \sigma C^{2} + \frac{5}{8} C^{2} \left( \log C^{2} - 3 \right) \Biggr] ,
           \label{V_h expression without z}
\end{eqnarray}
where the last term in (\ref{V_h expression with z}) yields a finite quantity, but it vanishes at $D=4$ after all, and thus it is removed here.

Substituting the expressions of $C$ into (\ref{V_h expression without z}), we separate it into the UV divergent part and the finite part at four dimensions by expanding around four dimensions. After subtracting the UV divergence by the counterterm, we obtain the one-loop contribution from the traceless tensor field as
\begin{eqnarray}
    V_{h} = \frac{5}{128} e^{4\sigma} \frac{t^{4}}{(4\pi)^{2}} M^{4} \left( \log \frac{t^{4}M^{4}}{16\mu^{4}} - \frac{21}{5} \right) .
      \label{result of V_h}
\end{eqnarray}
This also becomes independent of $\sigma$ apart from the overall factor.

\subsubsection{Physical cosmological constant}

Combining (\ref{result of V_varphi}) and (\ref{result of V_h}), we finally obtain the physical cosmological term, which is expressed as
\begin{eqnarray*}
    V= \Lambda e^{4\sigma} + V_{\phi} + V_{h} = v(\Lambda, M^{2},\alpha_{t},\mu) e^{4\sigma}  ,
\end{eqnarray*}
where
\begin{eqnarray*}
    v &=& \Lambda + \frac{\Lambda}{b_c} ( 7 - 2 \log 4\pi ) 
            - \frac{9\pi^{2} M^{4}}{2b_c^2} \left( \frac{25}{3} - 4 \log 4\pi \right) 
                \nonumber \\
      && - \left( \frac{\Lambda}{b_c} - \frac{9\pi^{2}M^{4}}{2b_c^2} \right) \log \frac{64\pi^{2}\Lambda}{b \mu^{4}}
            - \frac{6\pi M^{2}}{b_c} \sqrt{\frac{\Lambda}{b_c}- \frac{9\pi^{2}M^{4}}{4b_c^2}} 
              \arccos \left( \frac{3\pi M^{2}}{2\sqrt{b_c\Lambda}} \right) 
               \nonumber \\ 
      &&  + \frac{5}{128} \alpha_{t}^{2} M^{4} \left( \log \frac{(4\pi)^{2}\alpha_{t}^{2}M^{4}}{16\mu^{4}} - \frac{21}{5} \right) .
\end{eqnarray*}

Thus, we obtain the physical cosmological constant $v$ as a function of the renormalized quantities of the cosmological constant, the Planck mass and the coupling constant. What $v$ becomes independent of $\sigma$ reflects the invariance under the RG flow discussed before. The physical cosmological constant can be written in terms of $\tilde{\a}_t$, $\tilde{\Lam}$, and $\tilde{M}^2$ as $\tilde{v}(\lam)$ (\ref{running cosmological constant}). The approximation becomes good in the UV limit of $\lam \to 0$, where the running quantities decrease. The RG analysis indicates that the value of $\tilde{v}(\lam)$ fixed there is also preserved at low energies.

This result also indicates in the aspect of quantum field theories as follows. The cosmological constant in the action should be positive and not so small, since the action must be bounded from below significantly for the path integral to be stable. Nevertheless, we can take the physical cosmological constant to be any small value even though the cosmological constant in the action is not small.

\section{Conclusion and Discussion}
\setcounter{equation}{0}

We studied what the physical cosmological constant is in renormalizable quantum gravity indicating the asymptotic background freedom, which is formulated in a nonperturbative manner with treating the conformal-factor field exactly. We here examined the RG equation of the effective action with respect to the cosmological term that depends on the conformal-factor field only. We then found that, due to the diffeomorphism invariance and the nonrenormalization property of the conformal-factor field, the constant $v$ in the effective potential becomes invariant under the RG flow, and thus it gives the physical cosmological constant in this theory.

The physical cosmological constant was calculated at the one-loop level explicitly. It consists of the renormalized quantities of the cosmological constant, the Planck mass, and the coupling constant so that we can take its value to be small actually without suffering from the instability of the path integral as mentioned in the last part of the previous section. Since the theory is UV complete, it gives a different viewpoint on the cosmological constant problem usually discussed on the basis of the theory with a UV cutoff. The value of the physical cosmological constant will be passed on to the low-energy effective theory of gravity given by an expansion in derivatives of the metric field \cite{hhy06}.

What are the physical quantities in renormalizable quantum gravity which can be observed through cosmological experiments? The physical cosmological constant is one of them. The physical Planck mass also will be defined in the same way through the effective action whose form is fixed by diffeomorphism invariance. The dynamical IR scale $\Lam_{\rm QG}$ indicated from the dynamics of the traceless tensor field is also the  physical scale that is RG invariant.\footnote{ 
At the one-loop level, it is described as $\Lam_{\rm QG} =\mu e^{-1/\b_1 \a_t}$, where $\b_t = -\b_1 \a_t$.}
On the other hand, we cannot define the $S$ matrix as a physical quantity as usual, because spacetime still fully fluctuates even at $t=0$ so that there is no flat spacetime to defined the asymptotic state.

Cosmologically, the primordial power spectrum of the early Universe is one of the physical observables. In a linear approximation which becomes valid at large $b_c$, such a spectrum is given by the two-point function of the conformal-factor field, which provides the scale-invariant spectrum that is called the Harrison-Zel'dovich spectrum with positive amplitude $1/b_c$. In general, however, there is no systematical argument yet to derive observables or full spectra from Green functions among diffeomorphism-invariant operators. So, we cannot discuss the detail of the spectrum beyond the linear approximation at present. It is left as a future issue.

Finally, we mention the dynamics of spacetime in our quantum gravity theory. The background-free property will violate completely at the dynamical scale $\Lam_{\rm QG}$. Classical spacetime then emerges. The correlation length $\xi_\Lam =1/\Lam_{\rm QG}$ specifies the area in which quantum gravity is effective and outside of it becomes classical. It denotes that there is the ``minimal length" we can measure, without discretizing spacetime. Thus, spacetime is practically quantized with this scale. The existence of this scale has been indicated from the sharp falloff of the large angular component of the cosmic microwave background spectrum \cite{hy, hhy06, hhy10}.\footnote{ 
In this theory, inflation will ignite at the physical Planck mass scale, not the physical cosmological constant scale, and terminate at the dynamical scale $\Lam_{\rm QG}$ taking at the order of $10^{17}$ GeV lower than the Planck mass scale.}

\appendix

\section{Evaluation of Infinite Series}

The expression (\ref{V_varphi expression with z}) is here decomposed as
\begin{eqnarray*}
   V_{\phi} = \fr{1}{(4\pi)^2} \left( \fr{B}{2} - \fr{A^2}{4} \right) \left( - \fr{2}{D-4} - \gm + \log 4\pi \right) 
                  + U(z) .
\end{eqnarray*}
In the following, we consider $U(z)$, which is the part that depends on the IR mass scale $z$.

Let us consider the infinite series defined by
\begin{eqnarray}
    f(x,y) = \sum_{n=3}^\infty \sum^n_{m=0} \fr{n!}{(n-m)! m!} 
                \fr{(-1)^{n-1}(-1)^m}{2n(2n-m-1)(2n-m-2)} x^{2n-m} y^m .
         \label{series f}
\end{eqnarray}
Introducing the variables
\begin{eqnarray*}
    x^2 = \fr{B}{z^4},  \qquad xy = \fr{A}{z^2},
\end{eqnarray*}
the $z$-dependent part $U$ can be expressed as
\begin{eqnarray*}
   U(z) &=& \fr{1}{(4\pi)^2} \Biggl\{  -\fr{B}{2} \log z^2 + \fr{A^2}{4} \log z^2
           +\fr{AB}{4} \fr{1}{z^2} - \fr{B^2}{24} \fr{1}{z^4}
           + z^4 f \left( \fr{\sq{B}}{z^2}, \fr{A}{\sq{B}} \right) \Biggr\} .
\end{eqnarray*}

The infinite series $f$ is evaluated as follows. We first consider the function given by differentiating $f(x,y)/x$ twice with respect to $x$. It can be easily evaluated as
\begin{eqnarray*}
   \fr{\pd^2}{\pd x^2} \left\{ \fr{1}{x}f(x,y) \right\} 
    &=& \sum_{n=3}^\infty \fr{(-1)^{n-1}}{2n} \sum^n_{m=0} \fr{n!}{(n-m)! m!} x^{2n-m-3} (-y)^m 
          \nonumber \\
    &=& \sum_{n=3}^\infty \fr{(-1)^{n-1}}{2n} \fr{1}{x^3} (x^2 -xy)^n
          \nonumber \\
    &=& \fr{1}{2x^3} \left[ \log \left( 1+x^2-xy \right) -x^2 +xy + \half \left( x^2 -xy \right)^2 \right] .
\end{eqnarray*}
Then, integrating two times with respect to $x$, we obtain the following result:
\begin{eqnarray*}
    f(x,y) &=& x \int^x_0 du \int^u_0 dv \fr{\pd^2}{\pd v^2} \left( \fr{1}{v} f(v,y) \right) 
                  \nonumber \\
           &=& \fr{3}{4}x^2 +\fr{1}{24}x^4 + \fr{1}{4} \left( x-x^3 \right)y -\fr{3}{8} x^2 y^2
                  \nonumber \\
           &&  + \left\{ \fr{1}{4} \left( 1-x^2 \right) -\fr{1}{4}xy + \fr{1}{8}x^2 y^2 
                          \right\} \log \left( 1+x^2-xy \right)
                   \nonumber \\
           &&  + \left( \fr{x^2 y}{4} - \fr{x}{2} \right) \sq{4-y^2}
                    \arctan \left( \fr{x\sq{4-y^2}}{2-xy} \right)  .
\end{eqnarray*}
Indeed, this function reproduces the series (\ref{series f}) by expanding in $x$ and $y$.

Using this expression, we can obtain the loop correction to the effective potential by taking the vanishing limit of $z$ as
\begin{eqnarray*}
   \lim_{z\to 0} U(z)  = \fr{1}{(4\pi)^2} \left\{ \fr{1}{8} \left( 2B-A^2 \right) 
          \left( 3-\log B \right)    - \fr{A}{4} \sq{4B-A^2} \arccos \left( \fr{A}{2\sq{B}} \right)  \right\} .
\end{eqnarray*}
Here, we assume $A < 2\sq{B}$ and use the formula $\arctan(\sq{1-w^2}/w)=\arccos w$. This result can be extended to the range of $A > 2\sq{B}$ using the expression of the arccos function: $\arccos w = i\log(w+\sq{w^2-1})$ with $w >1$. If we take the limit of $B \to 0$, $U$ reduces to $A^2\{-3+\log A^2\}/8(4\pi)^2$.


\end{document}